\documentclass[journal]{IEEEtran}
\usepackage{gensymb}
\usepackage{graphicx}
\usepackage[caption=false,font=footnotesize]{subfig}
\usepackage{tikz}
\usepackage{cite}
\usepackage{amsmath}
\usepackage{array}
\usepackage{placeins}
\usepackage{hyperref}

\begin{document}

\title{Evaluating Positive Feedback Adiabatic Logic in 16\,nm FinFET with a Realistic Power-Clock}

\author{Franciszek~Łukowski, Maciej~Pyrzowski and Aida~Todri-Sanial \\ NanoComputing Research Lab, Integrated Circuits Group, Electrical Engineering Department \\ Eindhoven University of Technology, The Netherlands}

\maketitle 

\begin{abstract}
Adiabatic logic reuses the energy stored on load capacitances through quasi-reversible switching, enabling a lower minimum energy consumption than conventional static CMOS. Yet its practicality in FinFET technologies and at multi-GHz clock rates has yet to be investigated. This work provides a systematic evaluation of Positive Feedback Adiabatic Logic (PFAL) simulated in the TSMC 16\,nm FinFET process. A set of PFAL standard-cell gates were realised, along with two representative combinational circuits — a 2$\times$2 multiplier and a 4-bit comparator — and compared against static CMOS logic using the energy--delay product (EDP) and the energy advantage metric $\eta = E_{\mathrm{CMOS}} / E_{\mathrm{PFAL}}$. Transient simulations reveal three sources of non-adiabatic loss: two specific to the PMOS/NMOS latch, threshold-voltage-related loss and a previously unreported redundant charging of the output node and one related to the complexity of PFAL logic trees. The low-threshold Buffer/NOT cell achieves a minimum EDP of $1.23\times10^{-26}$\,J$\cdot$s at $V_{\mathrm{CLK}} = 0.6$\,V and $f_{\mathrm{CLK}} = 7.94$\,GHz, while PFAL preserves an energy benefit over static CMOS of up to roughly $5\times$ at reduced frequencies and elevated supply voltages. A parallel-coupled quadrature voltage-controlled oscillator is designed as a realistic four-phase power-clock generator. With this non-ideal supply, the Buffer/NOT energy stays within $2\,\%$ of the ideal sinusoidal case at $3$\,GHz. A loading study quantifies the phase shift and amplitude reduction induced by increasing fan-out. Overall, the results provide a design-oriented evaluation of PFAL in 16nm FinFET and a motivation to exploit adiabatic logic for future low-power system architectures.
\end{abstract}


\IEEEpeerreviewmaketitle

\section{Introduction} \label{sec:1}

\IEEEPARstart{E}{nergy} dissipation has become one of the defining constraints in modern digital integrated circuit design. As CMOS technology scales into the sub-22\,nm regime, static power due to subthreshold and gate-oxide leakage grows proportionally with device count, while dynamic dissipation accumulates, especially in high-throughput systems. Voltage scaling, once the primary lever for power reduction, is increasingly constrained by reliability margins, noise immunity, and the demands of high-frequency operation. The consequence is a per-operation switching-energy floor, set by the $\frac{1}{2}CV_{\mathrm{DD}}^2$ dissipation, that voltage scaling and circuit-level techniques can mitigate but not eliminate~\cite{rabaey2003}.

Adiabatic logic offers an alternative: rather than dissipating the energy stored on load capacitances as heat, it recycles that energy back into the power supply through quasi-reversible switching. The principle relies on two mechanisms~\cite{athas1994,teichmann2011}. First, output nodes are driven by a ramped power-clock waveform so that the voltage drop across any conducting transistor remains small throughout the charge transfer, minimising resistive dissipation. Second, the charge on the load is not dissipated to ground but returned to the oscillating supply during a subsequent recovery phase. The result is an energy floor that decreases with a longer ramp time, in contrast to the frequency-independent floor of static CMOS. 
Among quasi-adiabatic logic families, Positive Feedback Adiabatic Logic (PFAL)~\cite{vetuli1996} has demonstrated consistently robust energy efficiency at frequencies up to approximately 1\,GHz in planar CMOS~\cite{fischer2003,teichmann2011}. Systematic power-clock generation studies have established design foundations for four-phase supply architectures~\cite{jeanniot2018}.

Despite these advances, the viability of PFAL at deeply scaled FinFET technology nodes and at multi‑GHz operating frequencies remains largely unexplored. Existing studies have been limited to planar CMOS technologies with feature sizes of 45 nm and above. The present work therefore addresses the following research question: Do the energy‑recovery benefits of PFAL persist at advanced nodes; which additional or modified loss mechanisms arise in FinFET‑based implementations; under what conditions is functional correctness preserved; and are these benefits sustained when a realistic, non‑ideal power‑clock supply is employed? The working hypothesis is that PFAL maintains a net energy advantage over static CMOS at FinFET nodes for operating frequencies above 1 GHz, provided that the operating conditions remain within the correct‑operation envelope of the logic gates.

This paper presents a systematic characterisation of PFAL in the TSMC 16\,nm FinFET process. To address the question above, this work: (i) characterises a PFAL gate library and selected combinational circuits; (ii) maps the frequency–voltage envelope, where they remain functionally correct and energy-advantageous over CMOS logic; (iii) identifies the dominant loss mechanisms, and (iv) quantifies the energy impact of replacing an ideal power-clock with a fully transistor-level parallel-coupled voltage-controlled oscillator supply. 
The resulting characterisation maps the operating regimes, in which PFAL remains energy-advantageous over classical CMOS logic, 
providing a designer-oriented guide for integration of adiabatic logic in  
emerging low-power architectures.

\section{Adiabatic Switching Principle} \label{sec:2}
\begin{figure}%[h]
\centering
\includegraphics[width=0.5\columnwidth]{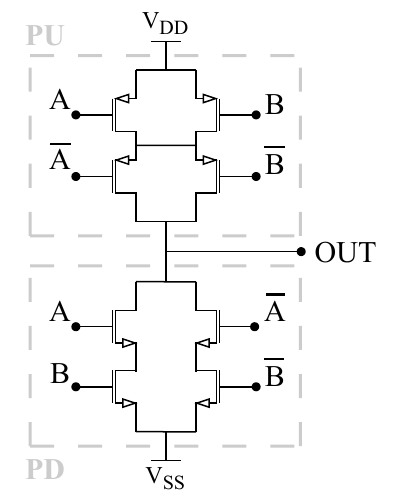}
\vspace{-10pt}
\caption{Circuit visualising the structure of a conventional static CMOS gate, implementing a XOR operation. NMOS transistors form a pull-down ($PD$) network, while PMOS transistors form a pull-up ($PU$) network. Based on the input combination the output $OUT$ is brought high ($\mathrm{V_{DD}}$) or low ($\mathrm{V_{SS}}$).}
\label{fig:general_CMOS_gate}
\vspace{-5pt}
\end{figure}
\subsection{Energy Dissipation in Adiabatic Charging}
% we can put the E_CMOS as a distinc Eq. if there is space
Conventional static CMOS logic gates, with an example visualised in Fig.~\ref{fig:general_CMOS_gate} based on a XOR operation, consist of a pull-up ($PU$) network, utilising PMOS transistors connected between the higher potential ($\mathrm{V_{DD}}$) and the output and a pull-down network, utilising NMOS transistors connected between the lower potential ($\mathrm{V_{SS}}$) and the output. Outcome is determined based on the input potentials ($A$, $B$ and their complements) by charging or discharging a load capacitance $C$. Charging of this load capacitance to a supply voltage $V_{\mathrm{DD}}$ through a switch with on-resistance $R$ dissipates a fixed energy~\cite{rabaey2003}:
\begin{equation}
    E_{\mathrm{CMOS}} =\tfrac{1}{2}\,C\,V_{\mathrm{DD}}^{2}.
\end{equation}
Adiabatic switching replaces the step-voltage source with a ramped supply of ramp time~$T$. The charging current $I = CV_{\mathrm{DD}}/T$ is constant, for a linear voltage ramp, given $T\gg RC$. The energy is $I^2RT=(RC/T)CV_{\mathrm{DD}}^2$ and can be further expressed as
\begin{equation}
  E_{\mathrm{adiabatic}} = \frac{RC}{T}\,C\,V_{\mathrm{DD}}^{2}
                         = \frac{2RC}{T}\,E_{\mathrm{CMOS}}.
  \label{eq:eadiabatic}
\end{equation}
From Eq.~\eqref{eq:eadiabatic} the dissipated energy is a fraction $\frac{2RC}{T}$ of $E_{\mathrm{CMOS}}$ and can in principle be lowered by increasing the ramp time~\cite{athas1994,teichmann2011}. 
\subsection{Adiabatic Logic Operation}
In addition to the slow-ramp condition, quasi-reversible operation requires two further constraints during switching, identified by the authors in~\cite{athas1994}: (i)~a transistor is not turned on while a voltage difference exists across its terminals, and (ii)~a transistor is not turned off while current flows through it. Any violation introduces a per-cycle, \emph{non-adiabatic} loss proportional to~$CV^{2}$, which, unlike $RC/T$, does not decrease with longer ramp time, restraining the achievable energy floor. Losses that obey the $RC/T$ scaling of Eq.~\eqref{eq:eadiabatic} are named \emph{adiabatic}~\cite{jeanniot2018}. 
\begin{figure}%[h]
\centering
\includegraphics[width=1\columnwidth]{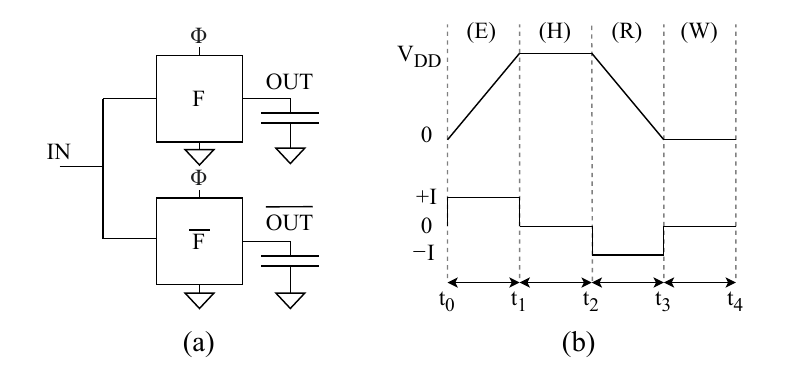}
\vspace{-20pt}
\caption{(a) General adiabatic gate and (b) example power-clock signal.}
\label{fig:general_info}
\vspace{-10pt}
\end{figure}
To realise adiabatic charging in a logic circuit, the constant supply $V_{\mathrm{DD}}$ is replaced by a periodic power-clock waveform~$\Phi(t)$. Fig.~\ref{fig:general_info} shows the general structure of an adiabatic gate and one period of the power-clock, which is divided into four phases: Evaluate~(E), Hold~(H), Recovery~(R), and Wait~(W). During the Recovery phase, the charge previously delivered to the output capacitance is returned to the supply. Throughout the work $f_{\mathrm{CLK}} = 1/(t_{4}-t_{0})$ denotes the power-clock frequency. Each gate requires both a logic function~$F$ and its complement~$\overline{F}$, producing differential outputs. Multi-stage pipelines require four power-clocks, each leading its predecessor by $90^{\circ}$, so that successive stages evaluate in sequence. 
The dual-rail structure of adiabatic logic ensures that one of the complementary outputs charges and recovers every cycle. The activity factor of an adiabatic gate, i.e., the fraction of clock cycles during which a node switches, is thus equal to 1~\cite{jeanniot2018}.

\section{PFAL Gate Library} \label{sec:3}

\begin{figure}%[h]
\centering
\includegraphics[width=0.93\columnwidth]{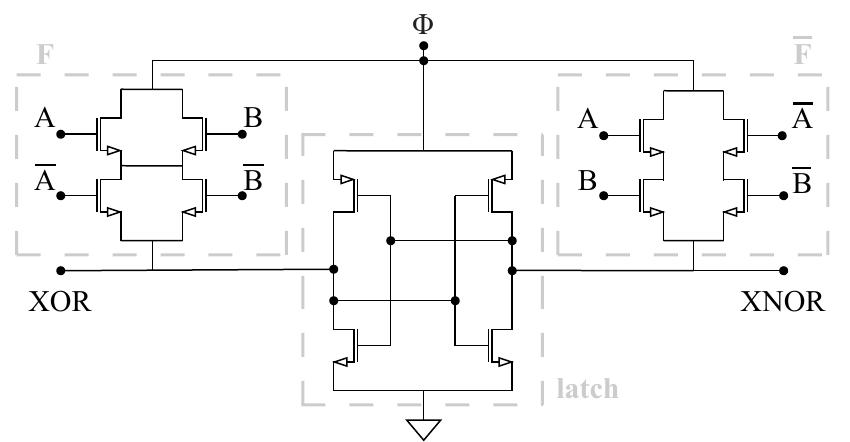}
\vspace{-10pt}
\caption{XOR/XNOR gate visualising the structure of a PFAL gate. The cross-coupled PMOS/NMOS latch is driven by a power-clock $\Phi$. Differential logic trees $F$ and $\overline{F}$ produce complementary outputs $OUT$ and $\overline{OUT}$, respectively.}
\label{fig:general_PFAL_gate}
\vspace{-5pt}
\end{figure}

Fig.~\ref{fig:general_PFAL_gate} exemplifies the PFAL gate structure based on a XOR/XNOR operation. The core of every PFAL gate is a pair of cross-coupled CMOS inverters, which form a positive feedback latch. Two complementary logic trees, $F$ and~$\overline{F}$, are placed in parallel with the PMOS latch transistors, each tree using only NMOS devices. One tree copies the pull-down network of the static-CMOS gate implementing the same function, while the other adopts the pull-up network topology by substituting PMOS with NMOS transistors. Since the trees are differential, PFAL gates additionally require complementary inputs and produce complementary outputs. This dual-rail structure can be exploited to reduce gate count in some combinational circuits, as demonstrated in Section~\ref{sec:5} with a 2$\times$2 Multiplier. Charge recovery in PFAL proceeds mainly through the latch PMOS devices, since both logic trees are disabled during the Recovery (R) phase. The latch PMOS is therefore the dominant energy-recycling path.

\begin{figure}%[h]
\centering
\includegraphics[width=\columnwidth]{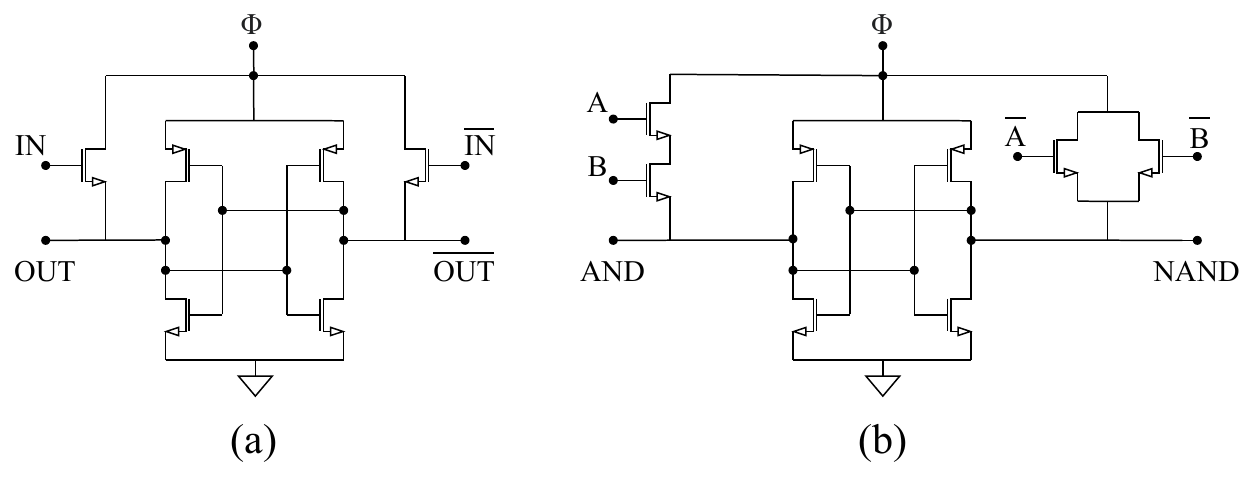}
\caption{PFAL (a) Buffer/NOT gate and (b) AND/NAND gate.}
\label{fig:xor_and_inv_pfal}
\end{figure}

The gate library characterised in this work further includes Buffer/NOT, AND/NAND and OR/NOR functions with circuit schematics visible in Fig.~\ref{fig:xor_and_inv_pfal}. The Buffer/NOT topology follows the original PFAL proposal of the authors in~\cite{vetuli1996}. The AND/NAND and XOR/XNOR gates were constructed by applying the same methodology as described previously. By De Morgan's laws, the AND/NAND topology realises OR/NOR operation when its inputs are swapped with their complements: the AND output then computes $\overline{A}\cdot \overline{B} = \overline{A+B}$ (NOR), while the NAND output computes $A+B$ (OR). A single schematic therefore implements both gate families, with the dual-rail outputs determining the function obtained at each terminal. Table~\ref{fig:andnand_pfal_truth_table} confirms this duality by inspection: reading the columns under $\overline{A}$ and $\overline{B}$ as inputs, the NAND output yields the OR function and the AND output yields NOR. Summary of each gate's transistor count is done in Table~\ref{tab:gate_summary}. Multi-input gate variants are obtained by extending the $F$ and~$\overline{F}$ trees in the same way as in static CMOS. 
\begin{table}%[h]
    \centering
    \caption{PFAL AND/NAND gate truth table.}
    \label{fig:andnand_pfal_truth_table}
    \begin{tabular}{c c c c c c}
    \hline
    \textbf{$A$} & \textbf{$B$} & \textbf{AND} & \textbf{NAND} & \textbf{$\overline{A}$} & \textbf{$\overline{B}$} \\
    \hline
    0 & 0 & 0 & 1 & 1 & 1 \\
    0 & 1 & 0 & 1 & 1 & 0 \\
    1 & 0 & 0 & 1 & 0 & 1 \\
    1 & 1 & 1 & 0 & 0 & 0 \\
    \hline
    \end{tabular}
\end{table}
\begin{table}%[h]
\centering
\caption{PFAL gate transistor count summary. All gates produce a valid output within one quarter of a power-clock cycle after the input is applied, corresponding to a per-gate computational latency of $T_{\mathrm{CLK}}/4$.}
\label{tab:gate_summary}
\begin{tabular}{l c c c}
\hline
\textbf{Gate} & \textbf{PMOS} & \textbf{NMOS} & \textbf{Total} \\
\hline
Buffer/NOT      & 2 & 4  & 6  \\
AND/NAND        & 2 & 6  & 8  \\
XOR/XNOR        & 2 & 10 & 12 \\
\hline
\end{tabular}
\end{table}
\section{Computing Losses in the 16\,nm Process} \label{sec:4}
Cadence transient simulations of the PFAL gate library identified three non-adiabatic loss mechanisms, one of which, to the author's knowledge, has not been previously reported in the PFAL literature. Two of these losses produce a per-cycle $CV^2$-type dissipation that does not vanish with longer ramp time, contrary to the $RC/T$-scaling adiabatic losses of Eq. \eqref{eq:eadiabatic}.  
Both originate in the PMOS/NMOS latch and are observed across the entire library. Below, they are explained using the Buffer/NOT circuit and its transient zoomed-in output waveform in Fig.~\ref{fig:zoomed_waveform}. The third loss is observed in circuits with more complex logic trees $F$ and $\overline{F}$ and is described based on the XOR/XNOR circuit and its transient waveform in Fig.~\ref{fig:xor_knee}.
\subsection{Redundant Output Node Charging}
In PFAL, the logic trees are placed in parallel with the PMOS latch transistors, between the power-clock~$\Phi$ and the output nodes (Fig.~\ref{fig:general_PFAL_gate}). During the Evaluate phase, the input signals are already settled (leading the clock by $90^{\circ}$), thus one logic tree is active and the other is blocked. Both PMOS transistors begin charging their respective output nodes as~$\Phi$ ramps up. The output node, whose logic tree is active, receives additional current through the parallel NMOS path and therefore charges faster than the complementary node, which is driven by its PMOS alone. When the faster-rising node reaches approximately $V_{\mathrm{th},n}$ of the cross-coupled latch NMOS, the corresponding latch transistor turns on and discharges the slower node to ground. However, by that point, the slower node has already been partially charged, and this energy is not recovered but dissipated through the NMOS to ground. The resulting redundant voltage peaks are visible in Fig.~\ref{fig:zoomed_waveform} at $t\approx 2.27$\,[ns] and $t\approx 2.6$\,[ns]. Their amplitude depends on the time required for the faster node to reach $V_{\mathrm{th},n}$. This time is set by the supply voltage, the output capacitance, and the conducting capability of the active logic tree. To the author's knowledge, the asymmetric partial charging of the complementary output node and its dissipation through the latch NMOS has not been explicitly identified as a distinct loss mechanism in the PFAL literature.
\subsection{Threshold Voltage Loss}
\begin{figure}%[h]
\centering
\includegraphics[width=0.85\columnwidth]{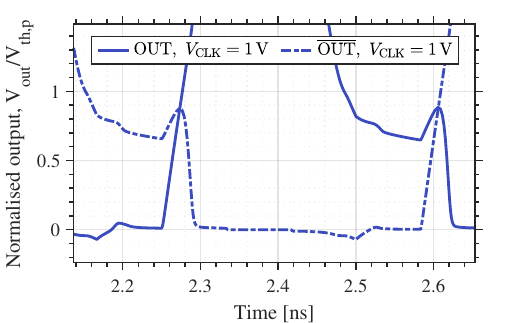}
\vspace{-5pt}
\caption{PFAL Buffer/NOT state transitioning output waveforms. Both traces show recovering output, with the threshold-voltage plateau starting to create at $V_{\mathrm{out}}\approx |V_{\mathrm{th},p}|$ and a redundant-charging peak preceding latch resolution.}
\label{fig:zoomed_waveform}
\vspace{-5pt}
\end{figure}

During the Recovery phase, both input signals are in the Wait state and neither logic tree conducts. The only available charge-recovery path is the PMOS latch transistor on the high-side output. Its conduction condition is
\begin{equation}
  V_{\mathrm{source}} - V_{\mathrm{gate}} \geq |V_{\mathrm{th},p}|,
  \label{eq:pmos_cond}
\end{equation}
where $V_{\mathrm{source}} = V_{\mathrm{CLK}}(t)$ and the gate is held at the complementary output, which is at ground. The PMOS therefore conducts only while $V_{\mathrm{CLK}}(t) \geq |V_{\mathrm{th},p}|$. Charge stored at voltages below this threshold cannot return to the supply and is discharged to ground through the latch NMOS in the following cycle, producing the plateau visible in Fig.~\ref{fig:zoomed_waveform} at $V_{\mathrm{out}} \approx |V_{\mathrm{th},p}|$, at $t\in(2.48; 2.62)$\,[ns]. This mechanism was previously reported in~\cite{jeanniot2018} and is confirmed in this work to persist at the TSMC 16\,nm FinFET node.

The threshold voltage loss motivates the use of low-threshold-voltage (LVT) devices in the gate library: a lower $|V_{\mathrm{th},p}|$ extends the recovery window, allowing more charge to return to the supply, while the correspondingly lower $V_{\mathrm{th},n}$ reduces the time the faster node needs to trigger the latch, thus decreasing the redundant charge on the complementary output. However, lower threshold voltage results in increased subthreshold leakage. At lower frequencies, the long cycle period accumulates more unrecoverable leakage charge per cycle, increasing the net per-cycle energy, even though the adiabatic losses are smaller~\cite{jeanniot2018}. Simulation of the Buffer/NOT gate (Section~\ref{sec:7}, Fig.~\ref{fig:16nmLVTvsSVT}) shows that LVT devices reduce energy consumption for operating frequencies above approximately 100\,MHz across the characterised supply range, while standard-threshold-voltage (SVT) devices are preferable below this region.
\subsection{XOR/XNOR Charge Redistribution}
In circuits with deeper logic trees such as the XOR/XNOR gate, the $F$ and~$\overline{F}$ logic trees contain intermediate nodes between series connected NMOS transistors that accumulate charge during the Evaluate (E) phase. During Recovery (R) phase, when the input-dependent recovery paths are not fully active, the stored charge on these intermediate nodes cannot be returned to the supply and is trapped. During the Wait phase, when inputs rise again, a path between trapped charge and the output may open (depends on the input combination) and the charge redistributes into the output node, extending the threshold voltage plateau and increasing the non-recoverable energy per cycle. 

\begin{figure}%[h]
\centering
\includegraphics[width=0.85\columnwidth]{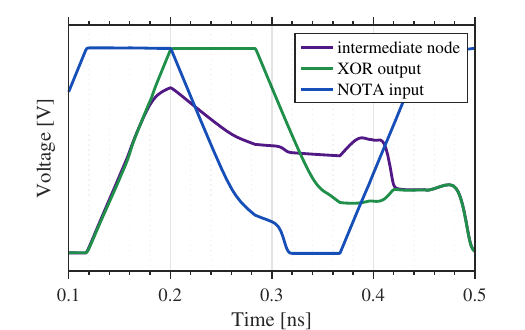}
\caption{XOR/XNOR gate node waveforms showing charge redistribution from the intermediate logic-tree node (purple) driven by input $NOTA$ (blue), into the XOR output (green) during Recovery. The resulting ``knee'' extends the non-recoverable voltage plateau beyond the $|V_{\mathrm{th},p}|$ level observed in the Buffer/NOT gate.}
\label{fig:xor_knee}
\end{figure}

Fig.~\ref{fig:xor_knee} illustrates this effect: after the output (green) reaches the $|V_{\mathrm{th},p}|$ plateau, input $NOTA$ (blue) rises high and charge from the intermediate node (purple) flows into the output, producing a visible ``knee'' that delays the final discharge. This additional loss mechanism, creates a trade-off between the complexity of a logic tree favoring speed (output resolves in $T_{\mathrm{CLK}}/4$), the efficiency of energy recycling and maximum operating speed. Furthermore, combined with the larger total capacitance of the 12-transistor XOR/XNOR gate, explains its reduced functional operating region compared to the Buffer/NOT and AND/NAND gates observed in the EDP characterisation of Section~\ref{sec:7}.

\section{Combinational Logic Circuits} \label{sec:5}
The gate library and loss analysis of the preceding sections establish the foundation for constructing more complex PFAL circuits.
Two combinational circuits were implemented and verified in TSMC 16\,nm: a 2$\times$2 multiplier and a 4-bit comparator. 

A design constraint distinguishes PFAL combinational circuits from their static CMOS counterparts. Since the gates are clocked, every stage must be supplied by the correct power-clock phase: each gate introduces a one-phase latency ($T_{\mathrm{CLK}}/4$), so multi-stage pipelines require careful phase assignment across all four clock phases. If a signal %must be
is reused in a later stage, a Buffer/NOT gate is inserted as a phase-alignment delay element, costing energy and area. 
All circuits were verified for functional correctness under trapezoidal, sinusoidal, and triangular power-clock waveforms. In all schematics, PFAL gate symbols adhere to the following convention: the upper two terminals carry $A$, $B$ and the lower two carry $\overline{A}$, $\overline{B}$. The output without a circle symbol is the positive result and the one with the circle is its complement.

\subsection{2$\times$2 Multiplier}

\begin{figure}%[h]
\centering
\includegraphics[width=0.7\columnwidth]{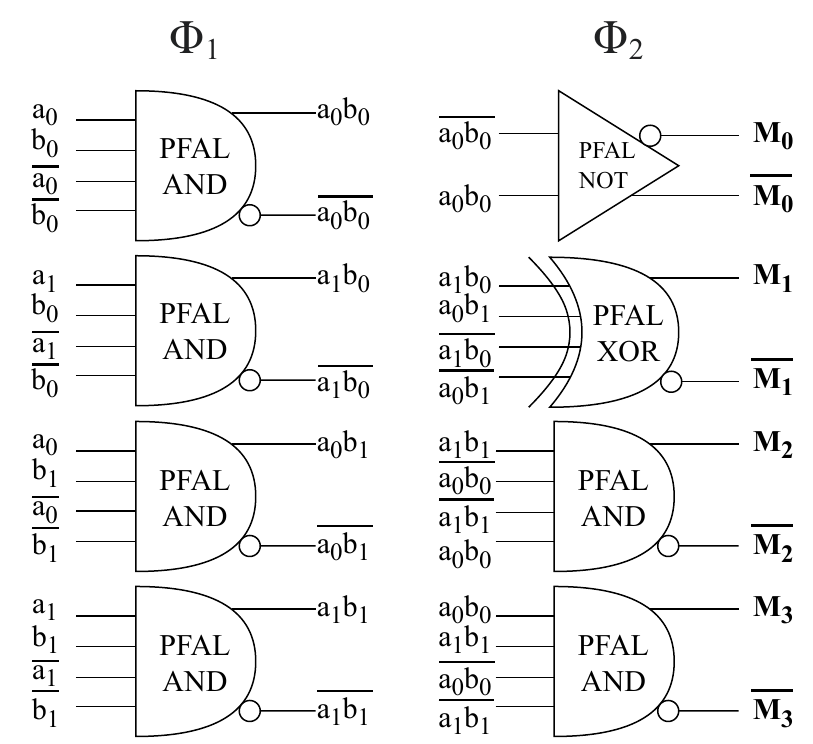}
\caption{2$\times$2 PFAL multiplier. $\Phi_1$: partial-product AND gates; $\Phi_2$: XOR, 2 ANDs and phase-alignment buffer.}
\label{fig:multiplierPFAL}
\end{figure}

A schematic of the PFAL 2$\times$2 multiplier is shown in Fig.~\ref{fig:multiplierPFAL}. The circuit architecture was adapted from~\cite{yousuf2015} for the PFAL topology used in this work and is organised in two power-clock phases. For 2-bit operands $A = a_1 a_0$ and $B = b_1 b_0$, the 4-bit product $M = M_3 M_2 M_1 M_0$ is computed as:
\begin{align}
  M_0 &= a_0 b_0, \label{eq:m0} \\
  M_1 &= (a_1 b_0) \oplus (a_0 b_1), \label{eq:m1} \\
  M_2 &= (a_1 b_1) \oplus (a_0 a_1 b_0 b_1) \Leftrightarrow (a_1 b_1) \cdot \overline{(a_0 b_0)}, \label{eq:m2} \\
  M_3 &= (a_0 b_0) \cdot (a_1 b_1). \label{eq:m3}
\end{align}
The key simplification over a standard CMOS implementation is in Eq.~\eqref{eq:m2}: the complemented partial product $\overline{a_0 b_0}$ is available directly from the complementary output of the first AND/NAND gate (Section~\ref{sec:3}), eliminating one XOR gate. The circuit uses 6~AND/NAND gates, 1~XOR/XNOR gate, and 1~Buffer/NOT gate for phase alignment of~$M_0$, overall 8~gates, 66~transistors (16 PMOS, 50 NMOS). The output is valid after two power-clock phases ($1 / (2f_{\mathrm{CLK}})$).

Functional correctness is verified against a reference truth table across all 16~input combinations and no mismatches were found.
\subsection{4-bit Comparator}
\begin{figure}%[h]
\centering
\includegraphics[width=0.75\columnwidth]{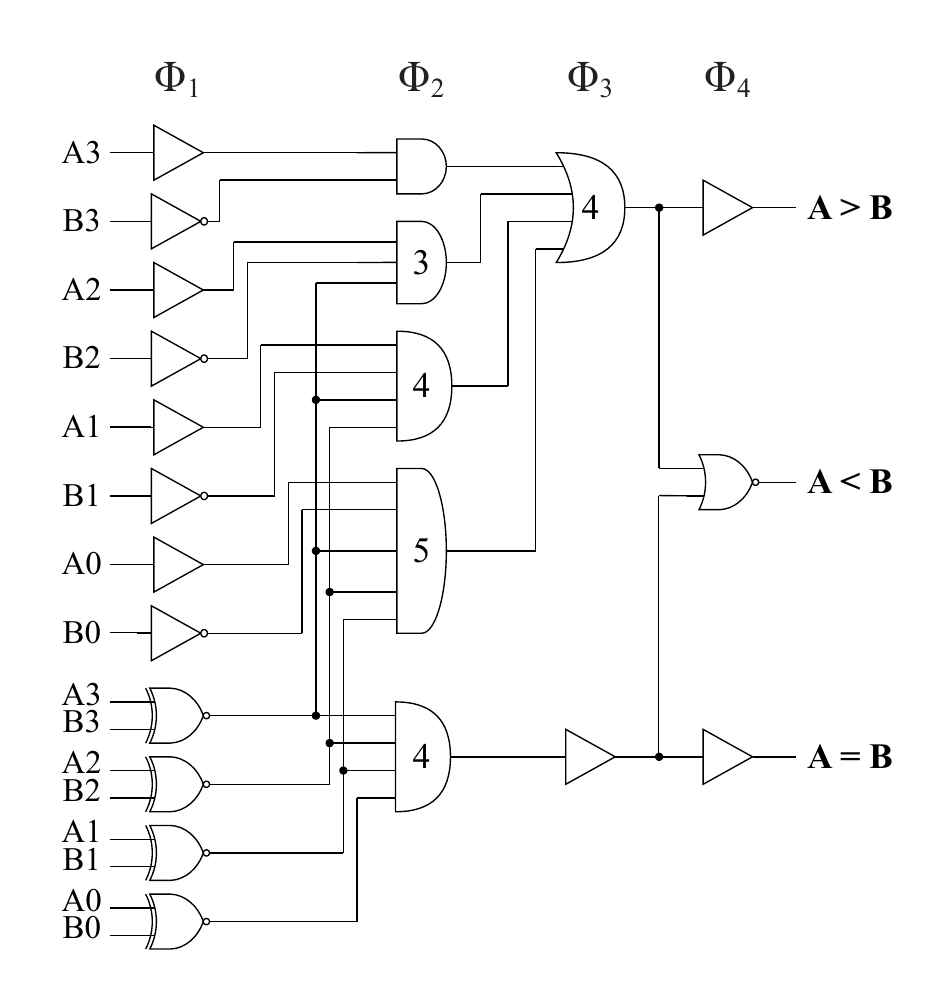}
\caption{Simplified schematic of a 4-bit PFAL comparator. Gate symbols show the gate function used. The complements are routed accordingly. Integer on a gate symbol indicates the number of inputs. Phase assignments are indicated by $\Phi_1$--$\Phi_4$ groupings.}
\label{fig:comp}
\end{figure}

The PFAL 4-bit comparator, shown in Fig.~\ref{fig:comp}, determines whether $A > B$, $A = B$, or $A < B$ for two unsigned 4-bit operands. The circuit topology was adapted from~\cite{alam2017} for the PFAL library described in Section~\ref{sec:3}, with input-polarity assignments, phase-alignment buffers, and per-phase gate placement specified by this work. In Fig.~\ref{fig:comp} gate symbols show the used gate function and the complements are routed accordingly. Integer on a gate indicates the number of inputs. No integer resembles a standard gate from the library of Section~\ref{sec:3}. The circuit is organised across four power-clock phases as follows:

\begin{itemize}
  \item $\Phi_1$: 8~Buffer/NOT gates for signal phase extension, and 4~XOR/XNOR gates to detect per-bit equality ($a_i \oplus b_i$).
  \item $\Phi_2$: 4~AND/NAND gates (2-, 3-, 4-, and 5-input) to form weighted magnitude-comparison terms, incorporating the bit-priority hierarchy, plus 1~AND/NAND gate (4-input) to combine previous equality terms and produce $A=B$ output.
  \item $\Phi_3$: 1~OR/NOR gate (4-input) to produce $A>B$ and 1~Buffer/NOT gate to extend preceding 4-input AND/NAND gate, thus $A=B$ output in phase.
  \item $\Phi_4$: 1~NOR/OR gate (2-input) to produce the final $A < B$ output, and 2 Buffer/NOT gates to extend $\Phi_3$ outputs for phase alignment.
\end{itemize}

\noindent The complete gate inventory is: 11~Buffer/NOT, 4~XOR/XNOR (2-input), 5~AND/NAND (1$\times$2-, 1$\times$3-, 2$\times$4-, 1$\times$5-input), and 2~OR/NOR (1$\times$2-, 1$\times$4-input), totalling 22~gates and 190~transistors (44~PMOS, 146~NMOS). Multi-input gates are constructed by extending the $F$ and $\overline{F}$ logic trees as described in Section~\ref{sec:3}. Since the design spans all four phases, the output is valid one full clock period $T_{\mathrm{CLK}}$ after the inputs are applied.

Functional correctness was verified using a set of 10 input pairs that test each decision path: both $A=B$ corners (all-zero and all-one operands), four $A>B$  cases in which each bit position in turn acts as the deciding bit, and the four $A<B$ symmetric cases obtained by swapping the operands. All tested combinations produced correct results. The architecture is scalable to $N$-bit comparators by extending the multi-input AND and OR gates. Although, the recycling efficiency was observed to decrease as the logic trees become more complex, due to higher path resistance and intermediate-node charge redistribution (Section~\ref{sec:4}).

\section{Power-Clock Implementation} \label{sec:6}
A real power-clock must store and release energy on every cycle while maintaining quadrature across four phases. An LC resonant tank naturally fulfills the energy-storage requirement, its sinusoidal output is simple to implement and sustain by introducing transconductance acting as compensation for internal tank losses.
A parallel-coupled quadrature voltage-controlled oscillator (P-QVCO), consisting of two NMOS-coupled, complementary cross-coupled LC oscillators, is therefore selected as the power-clock topology~\cite{andreani2002}. The NMOS coupling transistors enforce quadrature between the four output phases $Q{+}$, $Q{-}$, $I{+}$, $I{-}$, thus satisfying the four-phase clocking requirement of Section~\ref{sec:2}.
\subsection{Components and Quadrature Tuning}
\begin{figure}%[h]
\centering
\includegraphics[width=1\columnwidth]{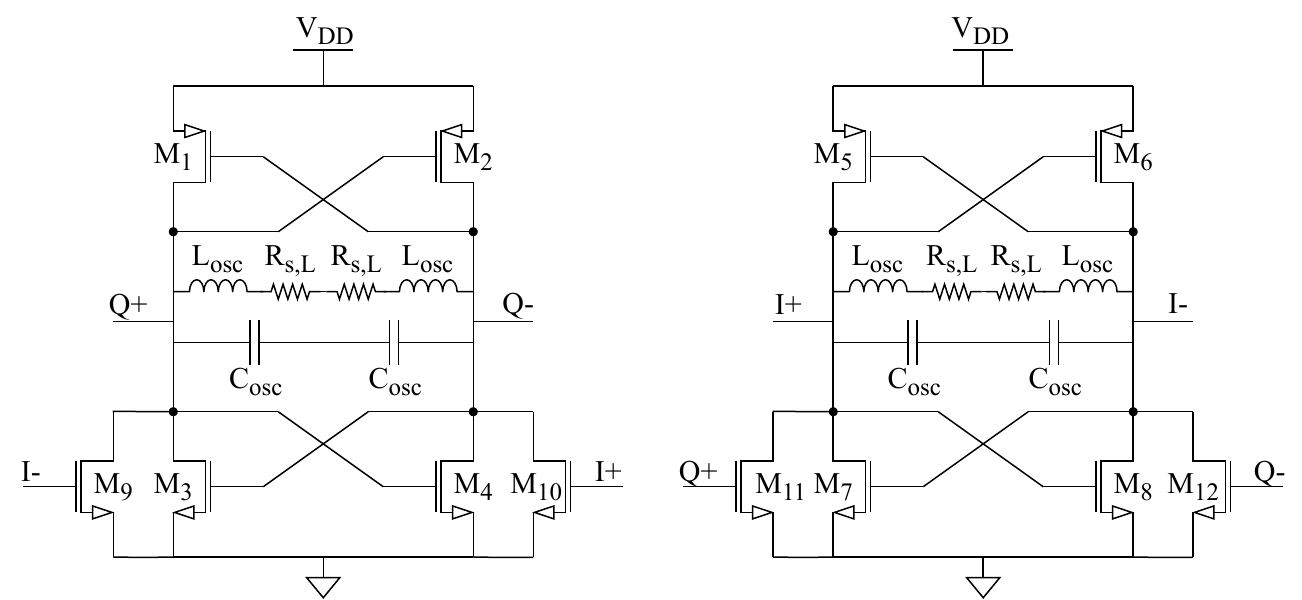}
\vspace{-20pt}
\caption{P-QVCO power-clock schematic. Two complementary cross-coupled LC cores are coupled through NMOS transistors $M_9$--$M_{12}$.}
\label{fig:qvco}
\vspace{-10pt}
\end{figure}

Fig.~\ref{fig:qvco} shows the oscillator schematic. Each LC core uses two off-chip inductors ($L = 2$\,nH, $Q = 20$), as the TSMC 16\,nm PDK does not include an inductor component. The required tank capacitance, determined by the target oscillation frequency $f_{\mathrm{osc}} \approx 3$\,GHz, was found to be $C_{\mathrm{osc}} \approx 1.41\,\text{pF}$ (Eq. \eqref{eq:cosc}).
The transistors were sized using the $g_m/I_d$ method~\cite{jespers_murmann_2017} targeting minimum energy consumption. The complete sizing procedure is documented in Appendix~\ref{app:qvco}. The coupling-width ratio $m = W_{\mathrm{coupling}} / W_{\mathrm{LC\,tank}}$ governs the trade-off between waveform purity and quadrature accuracy~\cite{andreani2002}, both important characteristics of adiabatic power-clocks. 

The NMOS coupling transistors ($M_9$--$M_{12}$ in Fig.~\ref{fig:qvco}) enforce quadrature between the two LC cores. In this work, $m= W_{\mathrm{coupling}} / W_{\mathrm{LC\,tank}}$ was swept from approximately $0.47$ to $1.2$. At $m \approx 0.47$ the simulated phase deviation from $90^{\circ}$ was approximately $15^{\circ}$ between the $Q$ and $I$ cores, increasing for lower values of $m$. Those values were not investigated further due to simulation time constraints.  Above $m \approx 1.2$ the sinusoid distortion became significant and the oscillator power increased significantly (clarified further in the section). The ratio was therefore treated as a design knob rather than fixed to a single value, allowing the energy measurements of Section~\ref{sec:7} to capture the sensitivity of PFAL efficiency to waveform shape and phase offset. The coupling ratio $m \approx 0.47$ was in the end selected as it yielded the lowest energy dissipation for the Buffer/NOT and AND/NAND gates in the measurements of Section~\ref{sec:7}, while still maintaining adequate phase differences for four-phase operation. Additionally, it provided a significantly lower overall power consumption of the oscillator comparing to the largest, tested $m$ (the largest, tested $m=1.2$ dissipated approx. 45\% more power). This is due to increase in width, of coupling transistors, by over a factor of 2, what significantly increases the oscillator capacitance and drawn current.

An NMOS-plus-PMOS coupling topology was also evaluated, while it produced a visibly cleaner sinusoid and better quadrature, the additional PMOS coupling transistors increased the oscillator power dissipation considerably (by 77\% at $m = 0.737$). Thus, the idea was not pursued.

Under no-load conditions, the oscillator settles at approximately $3$\,GHz, the output peak-to-peak voltage is $5$--$8$\,\% below the DC supply voltage, and the phase spacing is approximately $75^{\circ}$ between $Q$ and $I$ cores for $m=0.474$. The oscillator power consumption is $835$\,\textmu W at $V_{DD} = 800$\,mV and $1338$\,\textmu W at $V_{DD} = 900$\,mV both with $m=0.474$. The impact of the real power-clock supply on PFAL gate energy and the driving limits of the oscillator are quantified in Section~\ref{sec:7}.
\section{Results and Discussion} \label{sec:7}
\begin{figure}%[h]
\centering
\includegraphics[width=\columnwidth]{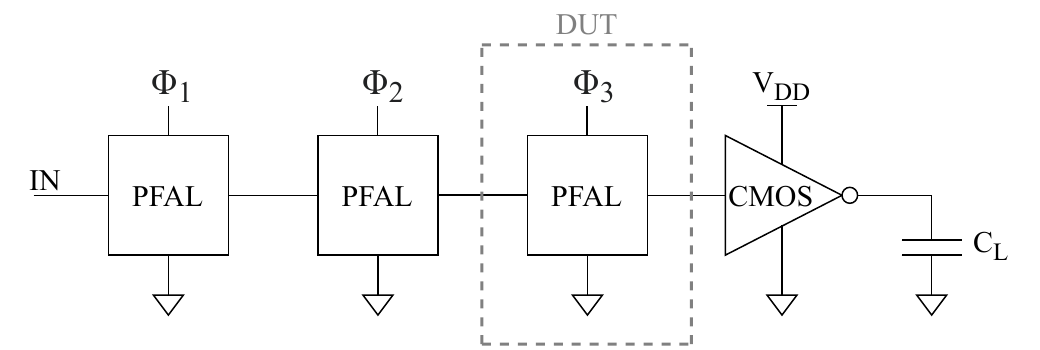}
\vspace{-20pt}
\caption{Testbench structure for PFAL gate and combinational
circuit characterisation. The DUT is preceded by two PFAL
buffers ($\Phi_1$, $\Phi_2$) and loaded with a minimum-sized
CMOS inverter and capacitor.}
\label{fig:testbench}
\vspace{-5pt}
\end{figure}

All gates and circuits presented in the preceding sections are implemented and simulated in Cadence Virtuoso using the TSMC 16\,nm FinFET (TSMC16ADFP) process design kits, with transistors based on the BSIM-CMG model~\cite{7313862}. Testbenches for each gate and circuit follow the structure shown in Fig.~\ref{fig:testbench}: the device under test (DUT) is preceded by two PFAL Buffer/NOT stages, each clocked by the next power-clock phase ($\Phi_1 \to \Phi_2 \to \Phi_3$), to provide a realistic adiabatic input waveform. The DUT output is loaded with a minimum-sized CMOS inverter and capacitor for the respective technology node. The energy drawn from the power-clock is recorded as:
\begin{equation}
  E = \int_{t_0}^{t_1} V_{\mathrm{CLK}}(t)\,
      I_{\mathrm{CLK}}(t)\,\mathrm{d}t,
  \label{eq:energy}
\end{equation}
where the integration window $[t_0,\,t_1]$ spans a complete set of all possible input combinations. The total energy is normalised by the number of input-combination sets, averaging out input-dependent variations arising during adiabatic operation, or by the number of clock periods.  
The figures of merit are: the energy-delay product $\mathrm{EDP} = E/f_{\mathrm{CLK}}$\,[J$\cdot$s], its minimum $\mathrm{EDP}_{\mathrm{min}}$, normalised $\mathrm{EDP}_{\mathrm{norm.}} = \mathrm{EDP}/\mathrm{EDP}_{\mathrm{min}}$\,[-], and the energy gain $\eta = E_{\mathrm{CMOS}}/E_{\mathrm{PFAL}}$, also called the Energy Saving Factor \cite{teichmann2011}.
The outputs are verified against reference logic values. An output was considered incorrect if it mismatched the reference output or if the residual distortion from Section~\ref{sec:4} propagated into the subsequent clock cycle with an amplitude exceeding $50$\,\% of $V_{\mathrm{CLK,max}}$, as such a level risks corrupting the following evaluation. For the 4-bit Comparator, the possible 256 input-output combinations made exhaustive simulation impractical. The circuit was tested against a representative set of combination pairs (Section~\ref{sec:5}). 
For the combinational circuits, the CMOS reference energy was estimated by summing the individually measured energies of the constituent static CMOS gates. This model neglects inter-gate loading effects or gate-specific load variations. Therefore, it approximates %a lower bound on 
the true CMOS energy consumption. All energy comparisons assume an activity factor of 1 for both PFAL and CMOS. In practice, lower CMOS activity factors would reduce the effective gain. In all EDP and energy-gain figures, the white region marks combinations where the gate fails the correctness criterion. Its boundary defines the functional operating limit. The white dots mark the exact simulated operating point. EDP is shown on a logarithmic scale and gain on a linear scale.
\subsection{Operating Boundaries and EDP Characterisation}
\begin{figure}%[h]
\centering
\includegraphics[width=\columnwidth]{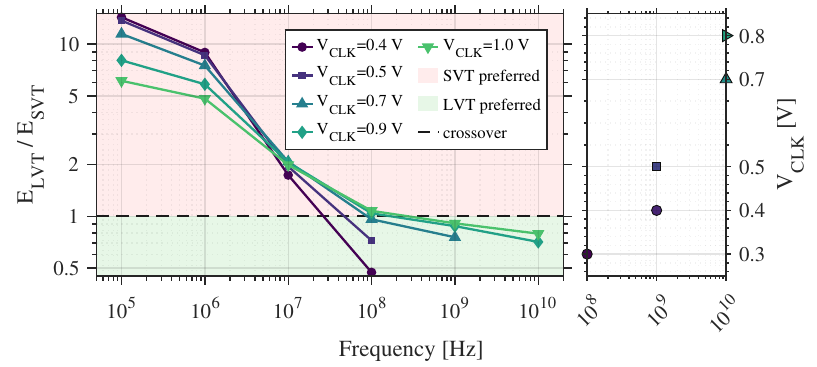}
\caption{Energy per period ratio of $E_{\mathrm{LVT}}/E_{\mathrm{SVT}}$ indicating preferred device type for a given operating point (left) and operating points where only LVT devices are functional (right) for different power-clock voltages $V_{\mathrm{CLK}}$.}
\label{fig:16nmLVTvsSVT}
\vspace{-10pt}
\end{figure}

\begin{figure}%[h]
\centering
\includegraphics[width=0.75\columnwidth]{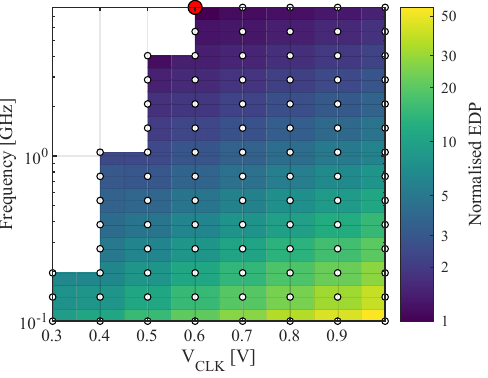}
\vspace{-10pt}
\caption{EDP per period heatmap of the LVT PFAL Buffer/NOT gate in 16\,nm with an ideal trapezoidal power-clock. The red dot marks the minimum EDP of $1.23\times10^{-26}$\,J$\cdot$s at $V_{\mathrm{CLK}} = 0.6$\,V and $f_{\mathrm{CLK}} = 7.94$\,GHz.}
\label{fig:Buffer/NOT_16nmEDP}
\end{figure}

Fig.~\ref{fig:16nmLVTvsSVT} compares the energy per clock period of LVT and SVT implementations of the PFAL Buffer/NOT gate across the characterised frequency and supply range. LVT devices extend the PMOS recovery window and reduce the latch-ambiguity time (Section~\ref{sec:4}), yielding lower energy consumption and a wider functional frequency range above approximately $10$--$100$\,MHz, depending on the supply voltage. 
Below this crossover, subthreshold leakage dominates and SVT devices become preferable \cite{jeanniot2018}. All subsequent results use LVT, consistent with the GHz target of this work.
 \\
Fig.~\ref{fig:Buffer/NOT_16nmEDP} presents the EDP heatmap of the LVT PFAL Buffer/NOT gate as a function of supply voltage and operating frequency. 
Within the correct operation envelope, the minimum EDP is consistently found at the lowest supply voltage corresponding to the highest achievable frequency. This is expected, as PFAL gate energy is robust against supply-voltage variations, while the $1/f_{\mathrm{CLK}}$ factor in the EDP decays with increasing frequency. Therefore, the frequency term dominates the product and drives the minimum toward the fastest feasible operating point. For lower frequencies, the heatmap shows that reducing the supply voltage maintains a comparable EDP level, providing a practical voltage-selection guide for a given frequency.

\begin{figure}%[h]
\centering
\includegraphics[width=0.75\columnwidth]{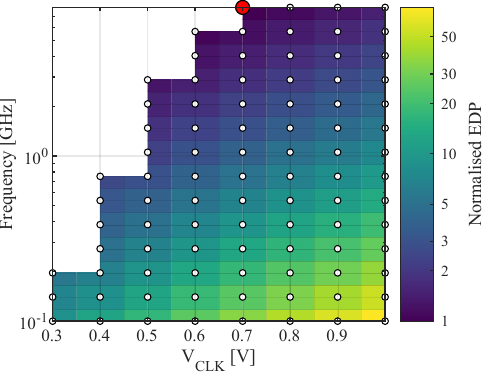}
\caption{EDP per input-combination set heatmap of the LVT PFAL AND/NAND gate in 16\,nm with an ideal trapezoidal power-clock. The red dot marks the minimum EDP of $5.75\times10^{-26}$\,J$\cdot$s at $V_{\mathrm{CLK}} = 0.7$\,V and $f_{\mathrm{CLK}} = 7.94$\,GHz.}
\label{fig:AND/NAND_16nmEDP}
\end{figure}

\begin{figure}%[h]
\centering
\includegraphics[width=0.75\columnwidth]{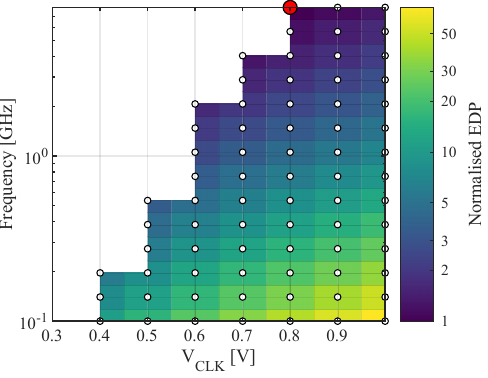}
\caption{EDP per input-combination set heatmap of the LVT PFAL XOR/XNOR gate in 16\,nm with an ideal trapezoidal power-clock. The red dot marks the minimum EDP of $1.01\times10^{-25}$\,J$\cdot$s at $V_{\mathrm{CLK}} = 0.8$\,V and $f_{\mathrm{CLK}} = 7.94$\,GHz.}
\label{fig:XOR/XNOR_16nmEDP}
\end{figure}

Fig.~\ref{fig:AND/NAND_16nmEDP},~\ref{fig:XOR/XNOR_16nmEDP} present the EDP heatmaps of the LVT PFAL AND/NAND and XOR/XNOR gates as a function of supply voltage and operating frequency respectively. Both heatmaps exhibit the same trend identified for the Buffer/NOT gate in Section~\ref{sec:7}: within the correct operation envelope, the minimum EDP is consistently found at the lowest supply voltage corresponding to the highest achievable frequency. Both plots, additionally, provide the frequency-voltage regimes where gates are operational. The XOR/XNOR gate has a smaller functional region than the Buffer/NOT, owing to its larger total capacitance (12 vs.\ 6~transistors per gate) and the intermediate-node charge redistribution described in Section~\ref{sec:4}. The AND/NAND gate falls between the two, consistent with its transistor count (8~transistors). Together with Fig.~\ref{fig:Buffer/NOT_16nmEDP}, these heatmaps complete the EDP characterisation of the gate library.

\subsection{16\,nm PFAL vs. CMOS with Ideal Supply}
\begin{figure}%[h]
    \centering
    \includegraphics[width=0.75\linewidth]{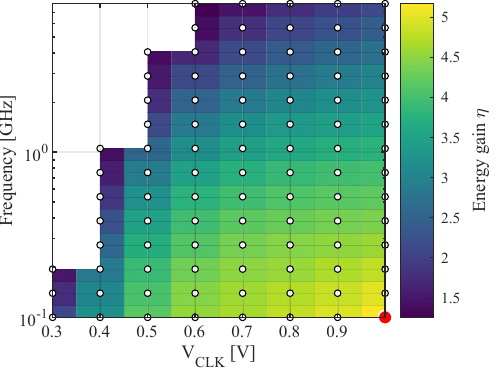}
    \vspace{-10pt}
    \caption{Energy gain per period $\eta$ map of the LVT PFAL Buffer/NOT gate in 16\,nm with an ideal trapezoidal power-clock. $\eta > 1$ indicates PFAL energy advantage. The red point marks the maximum gain $\eta \approx 5.2$ at $V_{\mathrm{CLK}}=1$\,V, $f_{\mathrm{CLK}}=100$\,MHz.}
    \label{fig:gain_inv}
    \vspace{-5pt}
\end{figure}

Fig. \ref{fig:gain_inv} presents the energy gain $\eta$ plot of the LVT PFAL Buffer/NOT gate over its static CMOS inverter equivalent at a given operating point. The PFAL circuits have energy advantage over CMOS across the whole available operating region. The energy gain has the highest magnitude for lower frequencies, due to the $1/T$ scaling as shown in Eq. \ref{eq:eadiabatic}, and for larger supply voltages, due to the fact that CMOS gates dissipate energy proportional to $V_{\mathrm{DD}}^2$ and are unable to recycle the charge.

\begin{figure}%[h]
    \centering
    \includegraphics[width=0.75\linewidth]{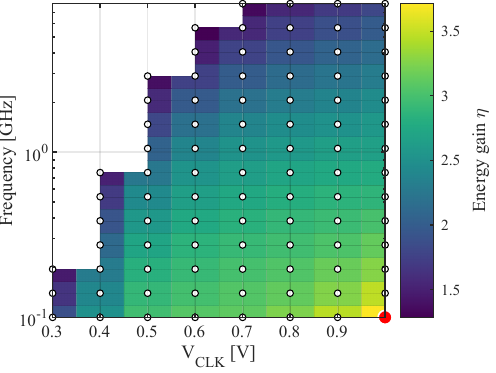}
    \caption{Energy gain per combination set $\eta$ map of the LVT PFAL AND/NAND gate in 16\,nm with an ideal trapezoidal power-clock. $\eta > 1$ indicates PFAL energy advantage. The red point marks the maximum gain $\eta \approx 3.7$ at $V_{\mathrm{CLK}}=1$\,V, $f_{\mathrm{CLK}}=100$\,MHz.}
    \label{fig:gain_nand}
\end{figure}

\begin{figure}%[h]
    \centering
    \includegraphics[width=0.75\linewidth]{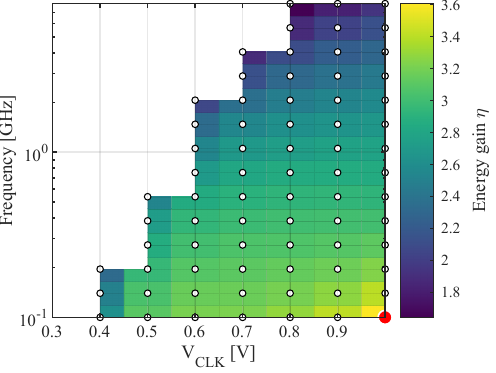}
    \caption{Energy gain per combination set $\eta$ map of the LVT PFAL XOR/XNOR gate in 16\,nm with an ideal trapezoidal power-clock. $\eta > 1$ indicates PFAL energy advantage. The red point marks the maximum gain $\eta \approx 3.6$ at $V_{\mathrm{CLK}}=1$\,V, $f_{\mathrm{CLK}}=100$\,MHz.}
    \label{fig:gain_xor}
\end{figure}

Fig.~\ref{fig:gain_nand} shows the energy gain $\eta$ of the LVT PFAL AND/NAND gate over its CMOS NAND equivalent, while Fig.~\ref{fig:gain_xor} presents the energy gain $\eta$ of the LVT PFAL XOR/XNOR gate over its CMOS XOR equivalent. The same trend observed for the Buffer/NOT gate holds: PFAL dissipates less energy than CMOS across the entire operating region. The energy gain has the highest magnitude for lower frequencies and for larger supply voltages. As expected, the operating region of the AND/NAND gate is smaller than that of the Buffer/NOT, and the XOR/XNOR is the smallest of the three, owing to their larger parasitic capacitances and the intermediate-node charging discussed in Section~\ref{sec:4}.

\begin{figure}%[h]
\centering
\includegraphics[width=\columnwidth]{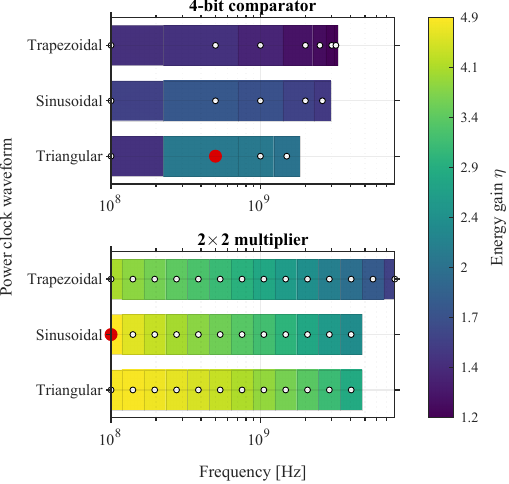}
\caption{Energy gain $\eta$ of the 2$\times$2 multiplier and 4-bit comparator realised in 16\,nm process supplied with $V_{\mathrm{CLK}} = 900$\,mV for trapezoidal, sinusoidal, and triangular power-clock waveforms. The red dots mark the maximal gain for each circuit. $\eta > 1$ indicates PFAL energy advantage.}
\label{fig:energygain_multcomp}
\end{figure}

Fig.~\ref{fig:energygain_multcomp} extends the analysis to the combinational circuits. The energy gain of the 2$\times$2 multiplier and 4-bit comparator is plotted as a function of frequency for $V_{\mathrm{CLK}} = 900$\,mV under three power-clock waveforms. Both circuits achieve $\eta > 1$ across the tested frequency range under all three waveform shapes, confirming that the gate-level energy advantage propagates to multi-gate systems. For the 4-bit comparator circuit, the largest gain $\eta \approx 2.1$ is found for triangular power-clock at $f_{\mathrm{CLK}} = 500$\,MHz, while for the 2$\times$2 multiplier, the largest gain $\eta \approx 4.9$ is found for sinusoidal power-clock at $f_{\mathrm{CLK}} = 100$\,MHz. The gain difference between the two discussed circuits results from larger losses in more complex logic trees of multi-input gates used in the comparator circuit. The CMOS reference energy for these circuits was estimated by summing the individually simulated CMOS gate energies, as described in Section~\ref{sec:7}. The reported gain therefore is an approximation.
\subsection{Real Power-Clock: Influence on Energy and Driving Limits} 
The PFAL Buffer/NOT gate was driven by the P-QVCO of Section~\ref{sec:6} and its energy compared against the ideal trapezoidal and sinusoidal references. Table~\ref{tab:real_supply} summarises the results at $f \approx 3$\,GHz for two supply voltages. The real sinusoidal supply reduces gate energy by $16$--$18$\,\% relative to the ideal trapezoid, and remains within $2$\,\% of the ideal sinusoid. At $900$\,mV, it is marginally more efficient than the ideal sinusoid, attributable to the $5$--$8$\,\% amplitude reduction (Section~\ref{sec:6}). The P-QVCO therefore does not significantly degrade adiabatic energy efficiency.

To broaden the assessment of the influence of a real power -clock on the energy recovery and to complete the gate library, the AND/NAND and XOR/XNOR gates were driven with the designed P-QVCO and their energy dissipation was measured. Table~\ref{tab:real_supply_nand} and Table~\ref{tab:real_supply_xor} summarise the results at $f\approx 3$\,GHz for two supply voltages $V_{DD}$. The AND/NAND gate follows the same trend as the Buffer/NOT: the real sinusoidal supply reduces energy by $26$--$28$\,\% relative to the ideal trapezoid and remains within $6$--$9$\,\% of the ideal sinusoid. The XOR/XNOR gate shows a slightly different behaviour: at $900$\,mV the real supply consumes approximately $6$\,\% more energy than the ideal sinusoid. Those small differences may come from the fact that this is an approximate mathematical model of the real system implemented and simulated in Cadence Virtuoso with ordinary differential equations. Additionally, the oscillator's peak-to-peak output voltages are $5$--$8$\,\% smaller than supply $V_{\mathrm{DD}}$ (Section~\ref{sec:6}). Nevertheless, these results further confirm that the P-QVCO does not significantly degrade adiabatic energy efficiency and that a sinusoidal power-clock is a practical replacement for the idealised waveforms.
\begin{table}%[h]
\centering
\caption{Energy ratio of 16\,nm LVT PFAL Buffer/NOT gate between: trapezoid, ideal and sinusoid, real (2nd column), \\ sinusoid, ideal and sinusoid, real (3rd column) power-clocks.}
\label{tab:real_supply}
\begin{tabular}{l c c}
\hline 
\textbf{Operating point}
  & $\dfrac{E_{\mathrm{real,sin}}}{E_{\mathrm{ideal,trap}}}$
  & $\dfrac{E_{\mathrm{real,sin}}}{E_{\mathrm{ideal,sin}}}$ \\[7pt]
\hline
$f=3$\,GHz, $V_{\mathrm{DD}}=800$\,mV & $0.836$ & $1.006$ \\
$f=3$\,GHz, $V_{\mathrm{DD}}=900$\,mV & $0.822$ & $0.9834$ \\
\hline
\end{tabular}
\vspace{-5pt}
\end{table} \\
\begin{table}%[h]
\centering
\caption{Energy ratio of 16\,nm LVT PFAL NAND/AND gate between: trapezoid, ideal and sinusoid, real (2nd column), \\ sinusoid, ideal and sinusoid, real (3rd column) power-clocks.}
\label{tab:real_supply_nand}
\begin{tabular}{l c c}
\hline 
\textbf{Operating point}
  & $\dfrac{E_{\mathrm{real,sin}}}{E_{\mathrm{ideal,trap}}}$
  & $\dfrac{E_{\mathrm{real,sin}}}{E_{\mathrm{ideal,sin}}}$ \\[7pt]
\hline
$f = 3$\,GHz, $V_{\mathrm{DD}}=800$\,mV & 0.723 & 0.908 \\
$f = 3$\,GHz, $V_{\mathrm{DD}}=900$\,mV & 0.741 & 0.936 \\
\hline
\end{tabular}
\end{table}
\begin{table}%[h]
\centering
\caption{Energy ratio of 16\,nm LVT PFAL XOR/XNOR gate between: trapezoid, ideal and sinusoid, real (2nd column), \\ sinusoid, ideal and sinusoid, real (3rd column) power-clocks.}
\label{tab:real_supply_xor}
\begin{tabular}{l c c}
\hline
\textbf{Operating point}  
  & $\dfrac{E_{\mathrm{real,sin}}}{E_{\mathrm{ideal,trap}}}$
  & $\dfrac{E_{\mathrm{real,sin}}}{E_{\mathrm{ideal,sin}}}$ \\[6pt]
\hline
$f = 3$\,GHz, $V_{\mathrm{DD}}=800$\,mV & 0.955 & 1.012 \\
$f = 3$\,GHz, $V_{\mathrm{DD}}=900$\,mV & 1.013 & 1.064 \\
\hline
\end{tabular}
\end{table}
\indent The driving capability of the P-QVCO is characterised by loading each oscillator phase with progressively more PFAL Buffer/NOT gate instances arranged in a one-dimensional pipeline, with a series interconnect resistance $R_{\mathrm{seg}} = 480$\,m$\Omega$ inserted between each consecutive gate, and measuring the resulting phase shift and amplitude loss relative to the unloaded oscillator output. The $480$\,m$\Omega$ value was adopted from the layout-extracted $65$\,nm model of~\cite{jeanniot2018} as a baseline. The $16$\,nm process was not laid out in this work. 
Table~\ref{tab:driving_limits} reports the results for $m = 0.474$. Up to $256$ gates per phase, the amplitude loss remains below $5$\,mV and the phase shift below $5.5^{\circ}$. However, at $384$ and $512$ gates, the phase shift exceeds $18^{\circ}$ and the amplitude drops by approximately $37$\,mV. The oscillator frequency falls to approximately $2.85$\,GHz for $512$ gate load. This $RC$ pipeline model provides an estimate of the driving capabilities of the designed Power Clock.
\begin{table}
\centering
\caption{P-QVCO driving limits for $R_{\mathrm{segment}} = 480$\,m$\ohm$, reported per generated phase w.r.t. the direct power-clock output sinusoid with $V_{\mathrm{CLK}}=900$\,mV, $f_{\mathrm{CLK}}\approx 3$\,GHz.}
\label{tab:driving_limits}
\begin{tabular}{l c c c c}
\hline
\textbf{Distortion type} & \textbf{192 gates} & \textbf{256 gates} & \textbf{384 gates} & \textbf{512 gates} \\
\hline
Phase shift & $5^{\circ}$ & $5.5^{\circ}$ & $18.5^{\circ}$ & $20.2^{\circ}$ \\
Amplitude loss & $4.5$\,mV & $5$\,mV & $36.5$\,mV & $37.5$\,mV \\
\hline
\end{tabular}
\vspace{-10pt}
\end{table}

\section{Conclusion}
\label{sec:8}

This work presented a systematic characterization of PFAL in the TSMC 16\,nm FinFET process. 
A small gate library such as Buffer/NOT, AND/NAND/OR/NOR, XOR/XNOR was implemented, together with a 2$\times$2 multiplier and a 4-bit comparator. Three non-adiabatic loss mechanisms were identified. Two specific to the PMOS/NMOS latch: threshold-voltage loss and redundant output-node charging. The latter is, to the author's knowledge, not previously reported in the PFAL literature. One related to the complexity of the logic trees of PFAL circuits -- intermediate node charge redistribution. The frequency--voltage operating boundaries of each gate were mapped through EDP heatmaps, with the LVT Buffer/NOT achieving a minimum EDP of $1.23\times10^{-26}$\,J$\cdot$s at $V_{\mathrm{CLK}} = 0.6$\,V and $f_{\mathrm{CLK}} = 7.94$\,GHz. Across the functional region, PFAL maintains an energy advantage over static CMOS of up to approximately $5\times$ at the most favorable operating points ($V_{\mathrm{CLK}} = 1$\,V and $f_{\mathrm{CLK}} = 100$\,MHz). Simulations of implemented combinational circuits showed propagation of energy recycling capabilities to multi-gate structures. A P-QVCO was designed to serve as a realistic power-clock source. Under this real supply, the Buffer/NOT gate energy remains within $2\,\%$ of the ideal sinusoidal reference and below the ideal trapezoidal energy at all measured operating points, confirming that energy recovery persists for real supplies. Load characterisation of the P-QVCO under a representative interconnect resistance shows the phase shift and amplitude attenuation increasing with fan-out, providing a quantitative basis for adiabatic-pipeline design.

Several limitations bounded the scope of these findings. The P-QVCO was designed around an off-chip inductor, as the 16\,nm PDK contains no inductor model. On-chip integration would require either a process with a sufficiently high quality factor or alternative topologies. The CMOS reference energy for the combinational circuits was estimated by summing the individually simulated gate energies. Full transient simulation of the equivalent CMOS circuits would tighten the comparison. Additionally, no layout-extracted parasitics were incorporated into any of the simulations. Natural extensions include a full ALU implementation and post-layout simulations using extracted parasitics. Further oscillator work could investigate the driving capabilities under physically routed interconnect rather than the model adopted here, optimize power consumption, and explore alternative tank topologies in which the gate capacitance itself replaces the dedicated tank capacitor, eliminating the need for $C_{\mathrm{osc}}$. Integration of an on-chip inductor presents an additional challenge. Taken together, these results confirm the stated hypothesis and demonstrate that PFAL remains a viable energy-recovery logic family at FinFET nodes and multi-GHz frequencies, even when driven by a transistor-level power-clock, establishing a quantitative foundation for the further investigation of adiabatic architectures for low-energy computing.

\bibliographystyle{IEEEtran}
\bibliography{IEEEabrv}

\appendices
\section{Power-Clock Design Process} \label{app:qvco}
This appendix documents the design procedure of the P-QVCO power-clock introduced in Section~\ref{sec:6}.

\subsection{Design Targets and Passive Components}

The oscillator was designed for a target frequency
$f_{\mathrm{osc}} = 3$\,GHz and a supply voltage of $900$\,mV, targeting a single-ended output swing of at least $800$\,mV to provide sufficient amplitude for the combinational circuits characterised in Section~\ref{sec:5} with headroom for more complex designs. A large tank capacitance is desirable to minimise the frequency shift caused by the capacitive load of driven PFAL gates.

A quality factor $Q=20$ and an inductance $L = 2$\,nH were assumed, as these are reasonable assumptions for off-chip inductors at GHz frequencies. The equivalent parallel resistance is then
\begin{equation}
  R_{p,L} = \omega\,L_{\mathrm{osc}}\,Q
           = 2\pi \cdot 3\times10^{9} \cdot 2\times10^{-9} \cdot 20
           \approx 754\,\Omega.
  \label{eq:rpl}
\end{equation}
The inductor series resistance $R_{\mathrm{s,L}}$ used for modelling, as can be seen in Fig.~\ref{fig:qvco}, is found as $R_{\mathrm{s,L}} = (\omega_{\mathrm{osc}}\cdot L_{\mathrm{osc}})^2/R_{\mathrm{p,L}}$. The corresponding tank capacitance follows from the LC resonance condition:
\begin{equation}
  C_{\mathrm{osc}} = \frac{1}{\omega_{\mathrm{osc}}^{2}\,L_{\mathrm{osc}}}
                    \approx 1.41\,\text{pF}.
  \label{eq:cosc}
\end{equation}
This capacitance is implemented as an NMOS-gate capacitor (NMOScap) available in the TSMC 16\,nm PDK, which captures parasitic effects more accurately than a lumped $RC$ model.

\subsection{Transistor Sizing}

The cross-coupled NMOS and PMOS transistors must supply a negative resistance sufficient to overcome the tank losses and sustain oscillation. The start-up condition requires a total effective transconductance~\cite{andreani2002}
\begin{equation}
  G_{m} \geq \frac{1}{R_{p,L}}.
  \label{eq:gm_startup}
\end{equation}
A factor-of-two margin was applied ($G_m \geq 2/R_{p,L} \approx 2.65$\,mS) to ensure headroom. The transconductance is split equally between the NMOS and PMOS devices ($g_{m,n} = g_{m,p} = G_m/2$). 
The $g_m/I_d$ methodology was used to determine the transistor widths~\cite{jespers_murmann_2017}. Technology-specific $g_m/I_d$, $g_{ds}/I_d$, $I_d/W$, and $C/W$ curves were extracted from the TSMC 16\,nm PDK at $V_{ds} = 0.2$\,V, a bias point modelling the lower-swing of the oscillating voltage magnitude, with a channel length $L = 20$\,nm. A $g_m/I_d$ operating point of $18$\,V$^{-1}$ was selected to minimise the oscillator power consumption. The required drain current and width for each device are
\begin{align}
  I_{d,\mathrm{req}} &= \frac{g_m}{(g_m / I_d)},
  \label{eq:idreq} \\[4pt]
  W_{\mathrm{req}} &= \frac{I_{d,\mathrm{req}}}{(I_d / W)}.
  \label{eq:wreq}
\end{align}
Table~\ref{tab:qvco_sizes} lists the transistor dimensions obtained from the $g_m/I_d$ sizing procedure. A width sweep was performed in Cadence Virtuoso to verify if smaller transistor dimensions sustain the oscillations. The analytically calculated values provided the best performance and were kept.

\begin{table}%[h]
\centering
\caption{P-QVCO transistor dimensions (TSMC 16\,nm).}
\label{tab:qvco_sizes}
\begin{tabular}{l c c}
\hline
\textbf{Device, quantity}            & $W$ [nm] & $L$ [nm] \\
\hline
NMOS (LC tank), $\times 4$ & 1460     & 20       \\
PMOS (LC tank), $\times 4$ & 1786     & 20       \\
NMOS (coupling), $\times 4$& 692      & 20       \\
\hline
\end{tabular}
\end{table}
\section{Extended Results} \label{app:extended_results}
This appendix provides extended results and supplementary
figures supporting the analysis of Section~\ref{sec:7}.
\subsection{16nm vs. 45nm Processes}
\begin{figure}
\centering
\includegraphics[width=\columnwidth]{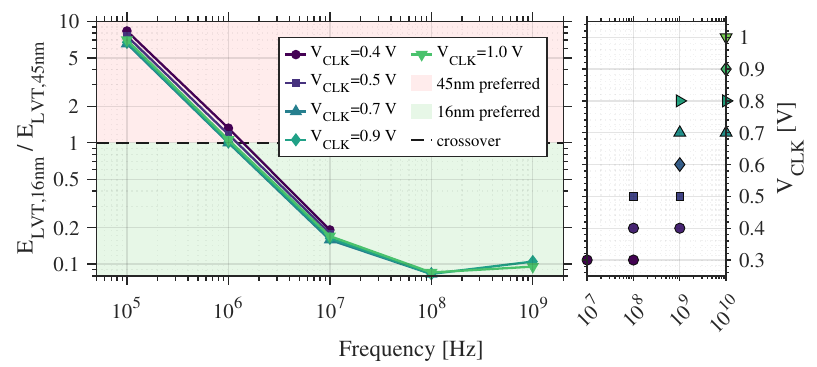}
\caption{Energy per period ratio of $E_{\mathrm{LVT,16nm}}/E_{\mathrm{LVT,45nm}}$ indicating preferred technology node for a given operating point (left) and operating points where only 16\,nm devices are functional (right) for different power-clock voltages $V_{\mathrm{CLK}}$.}
\label{fig:16nmvs45nm}
\end{figure}
Fig. \ref{fig:16nmvs45nm} illustrates the energy and speed comparison between $16$\,nm and $45$\,nm processes across different power-clock supply voltages. Below approximately $1$\,MHz, the $45$\,nm node dissipates less energy due to its higher threshold voltage and correspondingly lower subthreshold leakage. On the other hand, above this crossover the $16$\,nm node is more energy efficient, attributable to its lower threshold voltage permitting more effective charge recovery through the PMOS latch. Additionally, the $16$\,nm node sustains operation at GHz frequencies down to $V_{\mathrm{CLK}}= 0.4$\,V, whereas the $45$\,nm GPDK is limited to lower frequencies at comparable supplies.

\begin{figure}
\centering
\includegraphics[width=0.85\columnwidth]{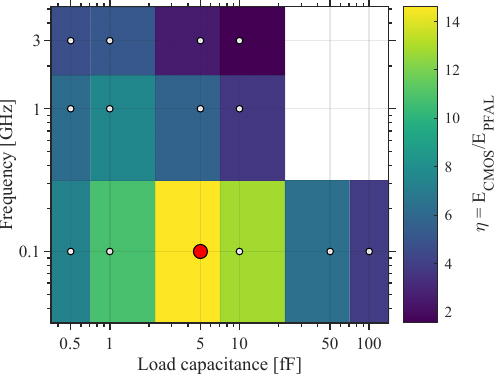}
\caption{Energy gain $\eta$ of the Inverter/Buffer gate realised in $16$\,nm for different load capacitances with supply voltage of $900$\,mV. $\eta > 1$ indicates PFAL energy advantage.}
\label{fig:loadsweepcmospfal}
\end{figure}
\begin{figure}
\centering
\includegraphics[width=0.85\columnwidth]{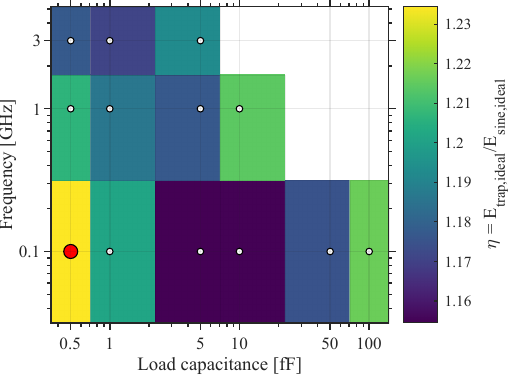}
\caption{Energy gain between sinusoidal and trapezoidal power-clock supplies of the Inverter/Buffer gate realised in $16$\,nm for different load capacitances with supply voltage of $900$\,mV. $\eta > 1$ indicates sinusoid energy advantage.}
\label{fig:loadsweeppowerclocks}
\end{figure}
\subsection{Load Capacitance Dependence}
\begin{table}
\centering
\caption{Energy ratio of 16\,nm LVT PFAL Inverter/Buffer gate between: trapezoid, ideal and sinusoid, real (2nd column), \\ sinusoid, ideal and sinusoid, real (3rd column) power-clocks for different load capacitances with $V_\mathrm{CLK}=900$\,mV.}
\label{tab:real_supply_loadsweep}
\begin{tabular}{l c c}
\hline
\textbf{Capacitance}  
  & $\dfrac{E_{\mathrm{real,sin}}}{E_{\mathrm{ideal,trap}}}$
  & $\dfrac{E_{\mathrm{real,sin}}}{E_{\mathrm{ideal,sin}}}$ \\[6pt]
\hline
$500$\,aF & 0.853 & 1.004 \\
$750$\,aF & 0.871 & 1.025 \\
$1$\,fF & 0.899 & 1.053\\
$2$\,fF & 0.952 & 1.132\\
$5$\,fF & 0.988 & 1.178\\
\hline
\end{tabular}
\end{table}
Minimum sized inverter and load capacitor of Fig.~\ref{fig:testbench} were replaced with an ideal capacitor, its capacitance value was swept and energy consumption per period was recorded for the Inverter/Buffer gate. In Fig.~\ref{fig:loadsweepcmospfal},~\ref{fig:loadsweeppowerclocks} the reported load capacitances reach a maximum of $100$\,fF, because for larger values the gates were not operational. The PFAL gate performs more energy efficiently across all tested operating points reaching a maximum energy gain $\eta\approx14$ for $f_\mathrm{CLK} = 100$\,MHz and $C_\mathrm{load}=5$\,fF. A sinusoidal power-clock performs between $16$\,$\%$ and $23$\,$\%$ better than the trapezoidal shape, matching the results from the waveform optimization. In Table~\ref{tab:real_supply_loadsweep} the maximum load value reaches $5$\,fF, as the real power-clock was not able to charge larger capacitances oscillating with $f_\mathrm{CLK}=3$\,GHz. It is found that the real sinusoidal clock performs more efficiently than the ideal trapezoidal power-clock for all tested operating points matching the results with the inverter load. For lower load values the real sinusoid performs comparably to an ideal sinusoid, however for larger capacitances the real source is less energy efficient reaching the maximum of approx. $17.8$\,$\%$ larger energy dissipation than its ideal equivalent.
\end{document}